\documentstyle[multicol,aps,prl]{revtex}

\begin{document}
\widetext

\draft
\title{Mixing by polymers: experimental test of decay regime of mixing.}
\author{T. Burghelea$^a$, E. Segre$^b$, and V. Steinberg$^a$}

\address{$^a$ Department of Physics of Complex Systems, $^b$ Department of Physical Services,\\
Weizmann Institute of Science, Rehovot, 76100 Israel}

\date{\today}
\maketitle
\begin{abstract}

By using high molecular weight fluorescent passive tracers with different diffusion coefficients and
by changing the fluid velocity we study dependence of a characteristic mixing length on the Peclet
number, $Pe$, which controls the mixing efficiency. The mixing length is found to be related to $Pe$
by a power law, $L_{mix}\propto Pe^{0.26\pm 0.01}$, and increases faster than expected for an
unbounded chaotic flow. Role of the boundaries in the mixing length abnormal growth is clarified. The
experimental findings are in a good quantitative agreement with the recent theoretical predictions.
\end{abstract}


\begin{multicols}{2}

\narrowtext

Mixing of a low diffusivity passive tracer by a turbulent flow, particularly in a region of small
scales, attracted recently significant attention from both theory\cite{1,2,3,4,5} and
experiment\cite{6,7}.  This so-called Batchelor regime of mixing exists in a viscous-convective range
of wave numbers, $k_K<k<k_B$, where the random-in-time velocity field can be treated as spatially 
homogeneous. Here
$k_K=(\epsilon/\nu^3)^{1/4}$ and $k_B=(\epsilon/\nu D^2)^{1/4}$ are the Kolmogorov dissipation and
the Batchelor wave numbers, respectively, $\nu$ and $D$ are the kinematic viscosity and the diffusion
 constant,
respectively, and $\epsilon$ is the rate of energy dissipation \cite{8}. The inverse injection scale
should be used here instead of $k_K$, if the former is larger than $k_K$. The range of the Batchelor
regime is defined by the value of the Schmidt number $Sc=\nu/D$, which can be rather large for a
viscous fluid with a low diffusivity tracer. The smoothness of the velocity field for developed
turbulence below the Kolmogorov scale is explained by an exponential decay of the velocity spectrum
there. The Batchelor-Kraichnan approach\cite{8} can be extended to an algebraic type velocity spectrum
$\propto k^{-\delta}$ with $\delta>3$. Then fluctuation spectrum of the velocity gradients scale as
$k^{-\delta+2}$, so that the flow becomes increasingly homogeneous on small scales, and the mixing
occurs mainly due to eddies corresponding to the border of the velocity smoothness. Batchelor and
Kraichnan\cite{8} considered a stationary regime of turbulent mixing with a tracer continuously
injected into a turbulent flow on a large scale. Then the mixing is produced by stretching and folding
of a tracer blob by the random velocity field creating components of higher and higher wave numbers up
to $k_B$. A more recent theory of the passive scalar decay in the Batchelor regime of mixing, which is
more relevant to a real experiment, makes predictions about the statistics of the passive
scalar\cite{5}.

Various theoretical predictions for the stationary and decay cases of the Batchelor regime of mixing
were verified recently in macro-systems \cite{6,7}. However, the theories mentioned \cite{8,2,5}
consider an unbounded case only. As shown just very recently \cite{9}, a non-uniform velocity
distribution particularly close to a wall, can alter significantly the efficiency of mixing in the
decay regime due to turbulent advection. Indeed, due to a reduced velocity near the wall the boundary
layer becomes a sink for the passive tracer. This excess of the tracer is intermittently injected from
the boundary layer into the bulk. Thus, the tracer decay, e.g. along a channel, is controlled by the
rate of tracer injection from the boundary layer that causes a significant slow down of the decay
compared with an unbounded case. As a result, dependence of the mixing length on the Peclet number,
$Pe=V_{av}d/D$, changes from logarithmic to algebraic, $L_{mix}\propto Pe^{1/4}$.
  Here $Pe$ defines
the relative rate of advection versus diffusive transport in a flow, $V_{av}$ is the average
longitudinal
 velocity, and $d$ is the channel width. Besides, the boundary layer width for
mixing in the decay regime in scales as $W\propto Pe^{-1/4}$. In order to verify these theoretical
predictions for scaling of the mixing length, of the width of the boundary layer, and of the correlation
function with $Pe$, we used a regime of elastic turbulence\cite{10}, which occurs at very low Reynolds
numbers, $Re\ll 1$, in a viscous fluid containing a small addition of high molecular weight
polymers, similar to the experiment reported in Ref.\cite{7}.\\
Solutions of flexible polymers are viscoelastic fluids \cite{11}. The distinctive characteristic of
such fluids is the presence of elastic stresses in a flow that relax with some macroscopic
characteristic time and increase non-linearly with the flow rate \cite{11}. The latter is the result
of the nonlinear mechanical relation between the elastic stresses and the rate of deformation of
polymer molecules in the flow. As shown theoretically and experimentally, the non-linear elasticity of
the polymer solution can lead to a purely elastic flow instability \cite{12,13}, and even to
elastically-driven turbulence as discovered recently  \cite{10}. These nonlinear elastic effects are
observable at an arbitrarily small $Re$, being practically independent of it, but strongly depend on
Weissenberg number, $Wi=\lambda \cdot \stackrel{\cdot }{\gamma }$. The latter reflects the competition
between non-linear elastic stresses and viscous stresses for the polymer part. Here $\lambda $ is the
polymer relaxation time, and $\stackrel{\cdot }{\gamma }$ is the shear rate. It was also shown
recently \cite{7} that a small addition of polymer to a viscous fluid can lead to a very efficient
mixing in a regime of the elastic turbulence generated in a curvilinear macro-channel, and that this
system happens to be an ideal test case for the Batchelor regime of
mixing.\\
In order to drastically reduce time for preparation of polymer solutions and at the same time to
reduce the amount of a fluorescent dye used, the experiment was carried out in a micro-fluidic set-up
based on a recently developed technology of soft lithography\cite{14,15}. Moreover, we demonstrated
just recently that the elastic turbulence and effective mixing can be obtained in a micro-channel\cite{16}
similar to those in a macro-channel\cite{7}.

 In this Letter we show that a micro-channel is the most appropriate
system to quantitatively test the recent theory of the Batchelor decay mixing regime in a bounded
system. As the result we were able to find the dependence of the mixing length on $Pe$,
the critical characteristic of a mixer.

The experiment was carried out in a curvilinear channel (inset in Fig.1) scaled down about 30 times,
 compared
to our macro-channel version \cite{7}. The channel was fabricated using a soft lithography \cite{15}
out of two components silicon elastomer, PDMS \cite{16}. As a working fluid we used a solution of 
80 ppm by weight
 high molecular weight polyacrylamide (M$_{w}$=18 Mda, supplied by
Polysciences). As a Newtonian solvent a $64.4\%$ sugar solution in a water with $1\%$ NaCl was used. The viscosity of the solvent $\eta_s=113.8$ mPas and the polymer solution
$\eta=137.6 $ mPas was measured at 22 $^\circ$C. The polymer relaxation time was $\lambda\approx 1.08$ sec,
and the solution density was $\rho=1.31$ g/l. The polymer solution was injected through two inlets of
the micro-channel, and 280 ppm of a fluorescein-labelled dextran was added in one of them. The
injection pressure was produced and regulated hydrostatically using vertical rails with rulers and
sliding stages. We have also checked in successive mixing experiments performed with the same solutions
 that the onset of the elastic instability remained unaltered, and the flow velocities were reproducible. 
 This suggests that neither the viscosity nor the characteristic relaxation time changed during 
 experiments.

One of the critical requirements of the experiment was to scan $Pe$ in as wide as possible range of
values. We varied $Pe$ about 40 times by using tracers with different $D$ and also by varying the flow
velocity. The former has been achieved by using various molecular weights fluorescein-conjugated
dextran (FITCD) that does not contribute to the elasticity of the solution, since it is highly branched 
and stiff. By varying the molecular weight of FITCD from 10Kda till 2Mda we were able to change $D$ of
the tracer more than 12 times, from $5.4\cdot 10^{-10}$ till $6.6\cdot 10^{-9}$ cm$^2$/sec. The values
of $D$ in water were taken from Ref.\cite{17} and then corrected according to viscosity of a solvent
used \cite{11}.

 In order to find a transition  to the elastic turbulence in a
micro-channel both components of the velocity (longitudinal and radial)were measured by microscopic
particle image velocimetry (micro-PIV) technique. We used the advantage of a narrow depth of a focus of
a high magnification objective (4$\mu$m) and an additional pre-filtering of the defocused image
features. We imaged 0.2 $\mu$m fluorescent particles with a CCD camera mounted on an epi-fluorescent
 inverted microscope and digitized with 12 bit, $640\times 512$ pixels resolution. As a result,
the velocity field was measured for each pressure drop with a spatial horizontal resolution of
$20\times 20$ $\mu$m and by using a standard cross-correlation algorithm between two consecutive
snapshots. An analog of the flow resistance, $\Delta p_s/V$, versus the pressure drop per segment,
$\Delta p_s$ is shown in Fig.1. Transition from a laminar regime to the elastic turbulence occurs
at $\Delta p_s\approx 50 Pa$. The average radial gradient of the longitudinal velocity, $\langle
\partial V/\partial r\rangle$ estimated from micro-PIV measurements, allows us to get the Weissenberg
 number at the transition, $Wi_{c}\approx 2.7 $. 

We studied the mixing dynamics in the elastic turbulence regime along the micro-channel by using a
confocal setup (Fluoview FV500 by Olympus) through an 40$\times$ infinity corrected objective with a
numerical aperture N.A.=0.85. Fluorescent light intensity profiles along a single line across the 
channel as a function of time were recorded by a 12 bit photomultiplier with a resolution 
of 0.18 $\mu $m/pixel at the acquisition rate of 56 lines/s.

 The interface separating the fluids with and without the dye was smooth  smooth and parallel to 
 the direction of the flow
in the case of a solvent injected, although little smeared by diffusion for $N>45$. Whereas in the
case of a polymer solution injected, the interface became unstable at $\Delta p_s\approx 50$ Pa.
First, the interface became unstable towards the end of the channel, and then in the range of the flow
rates used in the experiment on mixing, i.e. about twice of the onset value, the flow in the entire
channel became unstable and chaotic.

We investigated the evolution of the mixing patterns along the channel at various constant flow rates
above the transition to the elastic turbulence. The maximal longitudinal velocity 
was $V\approx 456 \mu m\cdot/sec$, and the radial velocity component was about 100$\mu$m\cite{16}. The 
corresponding average shear rate was obtained from the
average longitudinal velocity gradient,  $\dot{\gamma_{av}}=\langle \partial V/\partial r\rangle=3.7\ sec^{-1}$
that lead to $Re=\rho V_{av}d\eta^{-1}=1.6\cdot 10^{-4}$ and $Wi\approx 4.0$.

The smallness of $Re$ lead to a conclusion that the mixing observed was solely due to the elastic
turbulence similar to what occurred in a macro-channel\cite{7,10}, and so we expect to find  the
Batchelor regime of mixing like in a macro-channel\cite{7}. We varied $Pe$ more than 37 times from
$1.03\times 10^4$ up to $4.14\times 10^{5}$. As a result the Batchelor scale, $\eta_B=d\cdot
Pe^{-1/2}$, changed from $1\times 10^{-4}$ down to $0.15\times 10^{-4}$ cm that was of the same order
as a spatial resolution in the confocal measurements (better than $0.3 \mu m$).

As in a macro-channel the mixing in the micro-channel in the elastic turbulence regime turned to be
rather efficient as demonstrated in the space-time plots in Ref.\cite{16}. A quantitative measure of
the mixing efficiency is given by the i$^{th}$ order moments, $M_{i}$, of a probability distribution
function (PDF) of FITCD concentration, $c$,  defined by $M_{i}=\langle| c-\overline{c}|^{i} \rangle
/\overline{c}^{i}$, where $\overline{c}$ is the average concentration of FITCD. We present in Fig.2
the dependence of $M_1$ and $M_2$ on the position along the channel, $N$, for FITCD with molecular
weight 70 Kda (at $Pe=5.4\times 10^4$). In the inset of Fig.2 we show also the decay exponents of the
higher order moments, $\gamma_i$, as a function of their number, $i$, for five different $Pe$ values.
One notices that the saturation occurs for all curves at about the same value of $i=6$ but the
saturated value of the decay exponents depends on $Pe$. One can also notice from Fig.2 that, first,
both moments exhibit exponential decay up to $N\approx40$ as predicted by theory \cite{5,9}. Second,
the decay rate of $M_{2}$ is twice higher as of $M_{1}$. And third, $M_1$ is reduced more than 30
times and reaches the noise level at $N>40$, as well as $M_2$. Such measurements have been conducted
for six different values of $Pe$. The mixing length, measured in the number of units, $N_{mix}$,
 and obtained from the exponential decay of $M_{2}$,
is presented in Fig.3 as a function of $Pe$. The data are fitted by $L_{mix}\propto Pe^{0.26\pm 0.01}$
that is in a good agreement with the theory \cite{9}.

The latter result indicates that the efficiency of the turbulent mixing is reduced due to finite size
effects, which increase the mixing length in an bounded system compared with that
in an unbounded one. In order to establish the source of the algebraic dependence of $L_{mix}$ on $Pe$
instead of $ln(Pe)$, as predicted for an unbounded system, we studied the $Pe$ dependence of the
mixing layer width, $W$. The latter was identified from the spatial distribution across the channel of
the first and second moments of PDF of FITCD concentration fluctuations (see Fig.4). We associated
well-defined peaks in a spatial dependence of the moments with the boundary layer edge (note
asymmetry in left and right peak locations due to flow asymmetry). The average value of the right peak
location for the first moment, averaged along the channel as a function of $Pe$ is presented in the
inset of Fig.4 together with the fit: $W\propto Pe^{-0.28\pm 0.06}$. This result is once more in a
fair agreement with the theoretical prediction $W\propto Pe^{-1/4}$ \cite{9}.

And finally, the spatial correlation across the channel and temporal correlation functions of the
FITCD concentration fluctuations in the bulk and in the boundary layer were studied at different $Pe$. 
We found that the spatial correlation functions decay logarithmically at distances above
the diffusion length (that is varied from 1 to 4 $\mu$m) and are almost independent of $Pe$ (Fig.5). The
relatively narrow spatial range of the correlation function decay does not allow to reliably
distinguish between the logarithmic and the power-law behavior, if the exponent is small, as predicted
by the theory \cite{9} ( the exponent $\sim Pe^{-1/4}$).
 This closeness to the logarithmic decay can be explained by the fact that the mixing in the bulk
occurs rather efficiently. The mixing boundary layer permanently supplies the passive scalar into the
bulk that mimics the Batchelor stationary regime of mixing\cite{8} and leads to the logarithmic
decay\cite{18}. The temporal correlation functions in the bulk  and in the boundary 
layer (plots are not shown) show similar behavior with about twice larger correlation time in the boundary 
layer than in
the bulk ($t_{1/2}\approx 13$ sec instead of 6 sec in the bulk). On the other hand, using the Taylor
hypothesis we found that the correlation lengths along the the flow are rather close in the bulk and
the boundary layer ($d_{1/2}\approx 2.7$ mm vs 2.0 mm) and about 200 times larger than the correlation
length across the flow.

We are grateful to M. Chertkov and V. Lebedev for many useful and illuminating discussions, and to A.
Groisman for valuable remarks. One of us (T.B.) thanks  V. Kiss for his support in the confocal setup
measurements. This work is partially supported by Israel Science Foundation grant, by Binational
US-Israel Foundation grant, and by the Minerva Center for Nonlinear Physics of Complex Systems.

\begin{figure}
\caption { Flow resistance, $\Delta p_s/V$, vs. $\Delta p_s$. Arrow indicates the
onset of the elastic instability. The inset: Schematic diagram of 
a curvilinear micro-channel consisting of 64 base elements
smoothly interconnected. The number $N$ of the unit counted from the inlet is 
a natural linear coordinate along the channel.}
 \label{figa}
 \end{figure}

\begin{figure}
\caption { The first (squares) and second (dots) order moments decay vs. position, $N$, at
$Pe=5.4\times 10^4$; solid lines are the fit. The inset: The decay exponents of the moments, $\gamma_{i}$, as a function of the
moment number, $i$, at different $Pe$: diamonds- $Pe=1.03\times 10^4$, up-triangles- $Pe=2.0\times
10^4$, squares- $Pe=5.4\times 10^4$, circles- $Pe=1.3\times 10^5$, down-triangles- $Pe=2.5\times 10^5$.}
 \label{figb}
 \end{figure}

\begin{figure}
\caption {The mixing length, measured in the number of turns, $N_{mix}$, vs. $Pe$: squares are the data, 
solid line is the fit, and dotted line shows the logarithmic dependence.}
 \label{figc}
 \end{figure}

\begin{figure}
\caption{Spatial dependence of the first (gray) and second (black) order moments of FITCD
 concentration fluctuations across the micro-channel. The inset: Width of the boundary layer, $W$,
vs. $Pe$.Solid line is the fit.
}
 \label{figd}
 \end{figure}

\begin{figure}
\caption{ Correlation coefficients for the tracer concentration vs. $\Delta x/d$. Symbols are 
(1) $Pe=2.5\times 10^{5}$; (2) $Pe=1.0\times 10^5$; (3) 
 $Pe=5.4\times 10^4$; (4) $Pe=2.0\times 10^4$; (5) $Pe=1.03\times 10^4$.}
 \label{fige}
 \end{figure}

\end{multicols}
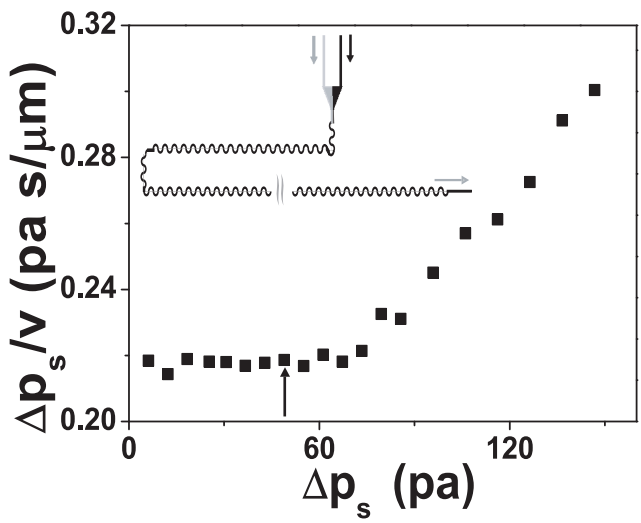
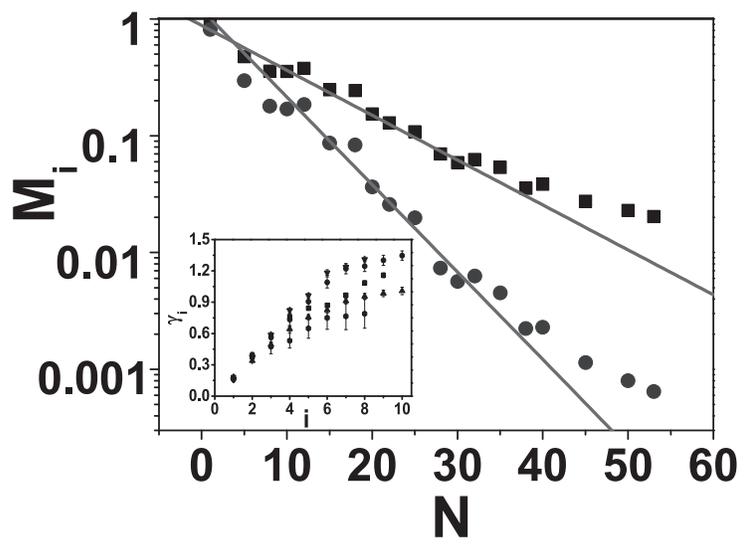
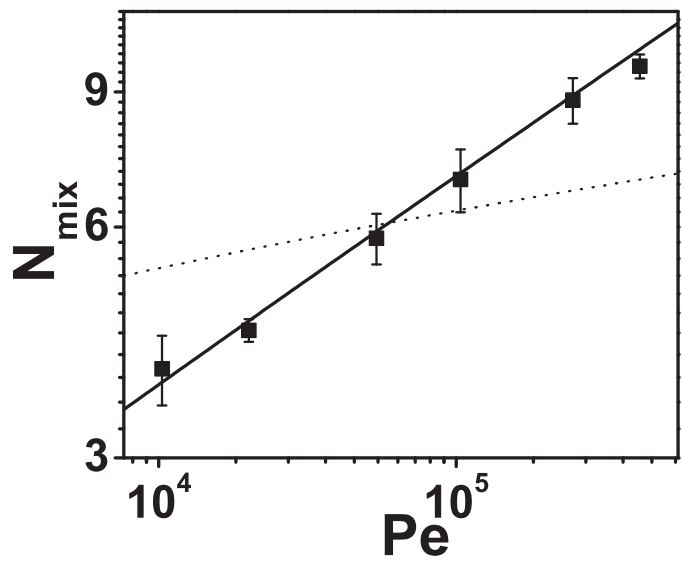
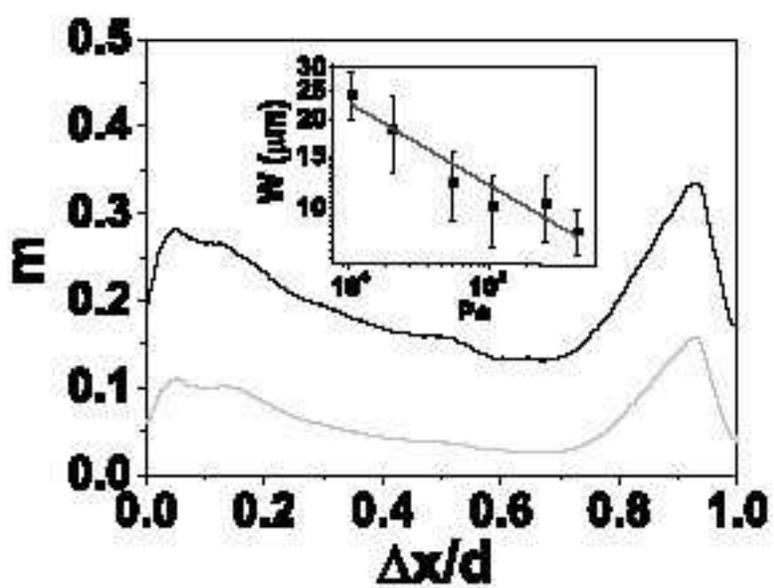
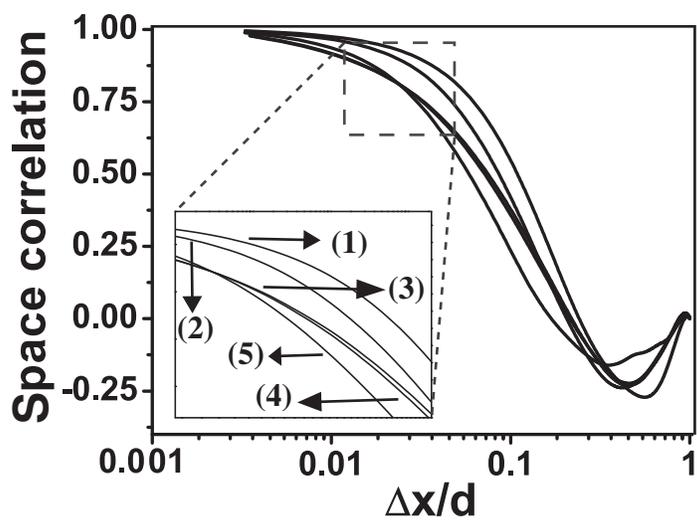
\end{document}